# Enhanced spatial modeling on linear networks using Gaussian Whittle-Matérn fields

Somnath Chaudhuri[a,b], Maria A. Barceló[a,b], Pablo Juan[a,c,*], Diego Varga[a,b,d], David Bolin[e], Håvard Rue[e], Marc Saez[a,b]

[a]*Research Group on Statistics, Econometrics and Health (GRECS), University of Girona, Spain*
[b]*CIBER of Epidemiology and Public Health (CIBERESP), Spain*
[c]*Department of Mathematics, University of Jaume I, Spain*
[d]*Department of Geography, University of Girona, Spain*
[e]*King Abdullah University of Science and Technology, Saudi Arabia*

**Abstract**

Spatial statistics is traditionally based on stationary models on $\mathbb{R}^d$ like Matérn fields. The adaptation of traditional spatial statistical methods, originally designed for stationary models in Euclidean spaces, to effectively model phenomena on linear networks such as stream systems and urban road networks is challenging. The current study aims to analyze the incidence of traffic accidents on road networks using three different methodologies and compare the model performance for each methodology.

Initially, we analyzed the application of spatial triangulation precisely on road networks instead of traditional continuous regions. However, this approach posed challenges in areas with complex boundaries, leading to the emergence of artificial spatial dependencies. To address this, we applied an alternative computational method to construct nonstationary barrier models. Finally, we explored a recently proposed class of Gaussian processes on compact metric graphs, the Whittle-Matérn fields, defined by a fractional SPDE on the metric graph. The latter fields are a natural extension of Gaussian fields with Matérn covariance functions on Euclidean domains to non-Euclidean metric graph settings.

A ten-year period (2010-2019) of daily traffic-accident records from Barcelona, Spain have been used to evaluate the three models referred above. While comparing model performance we observed that the Whittle-Matérn fields defined directly on the network outperformed the network triangulation and barrier models.

Due to their flexibility, the Whittle-Matérn fields can be applied to a wide range of environmental problems on linear networks such as spatio-temporal modeling of water contamination in stream networks or modeling air quality or accidents on urban road networks.

*Corresponding author: juan@uji.es

## 1. Background

Over the last few decades, advancements in computing and real-time data collection have enabled the collection of vast amounts of spatio-temporal data. As a result, statistical modeling of spatiotemporal data has gained more popularity and is now being utilized in various disciplines (Wood et al., 2004; Fuglstad & Castruccio, 2020). Applications range from the analysis of meteorological data, environmental data (Blangiardo et al., 2013), ecology (Zuur et al., 2017), and natural disasters such as forest fires (Juan et al., 2012; Serra et al., 2014), landslides (Lombardo et al., 2020), and earthquakes (Liu & Stein, 2016; Field et al., 2017). Additionally, spatiotemporal modeling is used in urban planning and strategic decision-making for issues such as traffic accidents (Prassanakumar et al., 2011; Liu & Sharma, 2018), criminal activities (Leong & Sung, 2015; Hossain et al., 2020), air pollution (Mota-Bertran et al., 2021, Saez & Barceló, 2022) and epidemiology and infectious disease dynamics (Schrödle et al., 2010; Moraga, 2020).

Depending on the objective of the study, various types of models are used with spatial and spatio-temporal data. With the development of Markov chain Monte Carlo (MCMC) simulation methods, researchers began to deal with these types of data using Bayesian methods (Gilks & Robert, 1996; Robert et al., 1999). To fit generalized linear mixed models (GLMM) in a spatial context, a Bayesian approach with MCMC simulation methods has traditionally been used. However, with the increase in data size and resolution, the computational burden of MCMC has become a critical issue (Rue et al., 2009; Rue et al., 2009; Taylor & Diggle 2014).

To address this issue, Rue et al. (2009), proposed a significantly faster solution as integrated nested Laplace approximations (INLA) which focuses on models that can be expressed as latent Gaussian Markov random fields (GMRF). Advancements in spatial statistics have made it possible to fit continuous spatial processes with a Matérn covariance function using INLA. Lindgren, Rue, and Lindström (2011) introduced a solution for the stochastic partial differential equation (SPDE) that provides a sparse representation of the solution fitting within the INLA framework. The solution for the spatial process can be represented as a sum of basis functions and associated coefficients, where the basis functions approximate the solution, and the coefficients follow a Gaussian distribution. This spatial model is implemented in INLA as the stochastic partial differential equation (SPDE) latent effect (Krainski et al., 2018). However, fitting this model with INLA requires the definition of a mesh over the study region to compute the approximation to the solution. Literature shows several research works where spatio-temporal models are constructed through Kronecker products of a spatial Matérn model and first- or second-order autoregressive models in time (Lindgren et al., 2015; Blangiardo & Cameletti, 2015; Bakka et al., 2018; Moraga, 2020; Lindgren et al., 2022) as applied within the framework of INLA.

### 1.1 Modeling on complex distributed spatial regions

Spatial models often assume isotropy and stationarity, implying that spatial dependence is direction invariant and uniform throughout the study area. However, these assumptions are violated when physical barriers are present in the form of geographical features as in case of complex island structures or in case of man-made barriers like disease control interventions and in animal species distribution problems. In these cases, the dependency among the observations should not be based on the shortest Euclidean distance between the locations but should take into account the effect of physical barriers and smooth across the land over holes or physical barriers (Bakka et al., 2019).

The traditional SPDE method triangulates the entire study area based on continuous geographic boundaries (Lindgren et al., 2011; Krainski et al., 2018). Problems arise in typical environmental research work such as modeling species distribution, where physical barriers such as mountains, roads or rivers could pose obstacles for the movement of species. Since propagation through those obstacles is not possible, spatial correlation should not follow the shortest path, but should travel around them. However, studies show that the meshes are usually generated for the entire study region, including the physical barriers. This approach involves generating an SPDE mesh for the entire study region, despite the presence of physical barriers that make the study area complex and distributed. For example, Lezama-Ochoa et al. (2020) used this approach to predict the occurrence of spine-tail devil ray species in the eastern Pacific Ocean. Bi et al. (2021) conducted a similar study to estimate seabird bycatch variations in the mid-Atlantic bight and northeast coast. In our review of the literature, we have explored several studies that have used the same approach to model complex land structures (Paradinas et al., 2015; Lourenço et al., 2017; R. De Jesus Crespo, 2019; Myer et al., 2019, Chaudhuri et al., 2023a). We aim to build upon these studies and further examine the application of INLA-SPDE in complex land structures, particularly in linear networks. Another serious concern to model observations in complex island structures is the anomaly related to the polygon structure of the coastlines. Coastlines are often considered as fractal structure, in the sense that any finite approximation will not be accurate (Bakka et al., 2019). For the same coastline polygons, different researchers may use varying approximations which can lead to conflicting interpretations and predictions. In that case, the model loses its scientific credibility. It is worth mentioning that a stationary model cannot be aware of the coastline structure and will inappropriately smooth over the features. In spatial modeling, classical models become unrealistic when they fail to account for holes or physical barriers in the landscape. This can lead to further unrealistic assumptions.

Bakka et al. (2019) introduced the barrier model as a solution to the limitations of existing models. Unlike traditional models, this new model does not rely on the shortest distance around a physical barrier or specific boundary conditions. Instead, it provides a non-stationary Gaussian random field which can handle sparse data and complex barrier structures. Additionally, the computational cost is comparable to that of a stationary model. In a recent study, Martínez-Minaya et al. (2019) have used the barrier model approach to design a Bayesian hierarchical species distribution model (SDM) to determine vulnerable habitats for bottlenose dolphins in the Northern Sardinia archipelago in Italy. Likewise, the use of barriers is also crucial in the control of infectious diseases. Cendoya et al. (2022) studied the impact of barriers on the spatial distribution of a quarantine plant pathogenic bacterium in Alicante, Spain. The simulation study in the Archipelago (Bakka et al., 2019) and other applications demonstrate that while barrier models have a similar computing cost to their corresponding stationary models, they are more flexible and realistic when used in complex spatial regions with physical barriers. However, some anomalies are found when the barriers are infinitely thin, in those cases artificially thicker barriers, such that the width is at least a mesh triangle, can make the model functional. Li et al. (2023) extended the barrier model by introducing a multi-barrier model that can characterize areas with different types of obstacles or physical barriers. Authors compared the stationary Gaussian model, barrier model, and proposed multi-barrier model using real burglary data, and the results suggest that all three models have similar performance.

**1.2 Modeling on metric graphs**

In many environmental applications such as urban road networks it is essential to define statistical models on linear networks. A major focus of research in this field has been spatiotemporal modeling of traffic accidents on urban road networks (Karaganis & Mimis, 2006; Castro et al., 2012; Boulieri et al., 2017; Liu et al., 2019). Studies like, Xu & Huang (2015), Wang et al. (2019) and Eboli et al. (2020), effectively capture the spatial dependence and heterogeneity in traffic accident data, improving the accuracy and robustness of predictions compared to traditional regression models on road networks. Recently, a number of models on road safety have been proposed following Bayesian methodology. Cantilo et al. (2016) used a combined GIS-Empirical Bayesian approach in modeling traffic accidents in the urban roads of Columbia. A similar research work on urban road network of Florida by Zeng & Huang (2014) explored Bayesian spatial joint modeling of traffic crashes. A space-time multivariate Bayesian model was designed by Boulieri et al. (2017) to analyze road traffic accidents by severity in different cities of UK. Recently, Galgamuwa et al. (2019) used Bayesian spatial modeling using INLA in predicting road traffic accidents based on unmeasured information at road segment levels. Due to densely distributed nature of the road segments, the majority of these studies used continuous spatial structures and traditional spatial stationary models such as Matérn fields. As a result, though the sampling points (here the traffic accident locations) are mainly located on the road networks, the SPDE triangulations are designed on the entire study area, including the areas without road network. Thus, the model result might be unpreventably generalized as it is going to estimate predicted values for the regions where there is no chance of incident to occur. In this context, Chaudhuri et al. (2022) recently proposed spatiotemporal modeling of road traffic accidents using explicit network triangulation on the road network of London, UK. In a similar study by Chaudhuri et al. (2023b), SPDE triangulation has been designed precisely on linear road networks of Barcelona, Spain to generate dynamic traffic accident risk maps. The methodology used in these two studies is a novel approach to perform spatio-temporal analysis precisely on road network and contributes to the relatively small amount of literature in this domain. However, in both cases, the complex boundary regions of the buffer road network result in high boundary effect, which can influence the spatial effects of the models. This is a serious limitation of the SPDE network triangulation approach. In general, Gaussian random fields with Matérn covariance functions are a popular choice but they have a stationary and isotropic covariance structure. Dawkins et al. (2021) made a novel attempt to apply a barrier model on linear road networks. This research showed a non-stationary approach to accurately estimate air quality levels on the roads of Brisbane, Australia using Bayesian methods. The study accounted for the topographical diversity of buildings in proximity to city roads by employing a non-stationary barrier model that extends upon the INLA framework.

In recent years, there has been a growing interest in examining point patterns on linear networks. In this context, Møller & Rasmussen (2022) proposed models for isotropic Gaussian processes and various Cox processes with isotropic pair correlation functions on linear networks, employing isotropic covariance functions based on geodesic or resistance metrics. On a different note, Porcu et al. (2023) conducted a recent study focusing on graphs with Euclidean edges. They introduced stationary non-separable space-time covariance functions to model spatio-temporal data on generalized networks, such as Euclidean trees and linear networks. Filosi et al. (2023) explores generalized networks with evolving topological structures, including graphs with Euclidean edges, addressing linear and periodic structured circular time scenarios and demonstrates the construction of proper semi-metrics for these temporally changing networks.

On the other hand, Bolin et al. (2022) presented an alternative to use the Euclidean distance by defining similar models with a non-Euclidean metric on the network. Because it can be challenging to find a class of positive definite functions suitable for creating Gaussian fields on metric graphs when using a non-Euclidean metric, Bolin et al. (2022) proposed characterizing the field through a SPDE defined directly on the network. The resulting model the Whittle-Matérn fields, can be used in inference without any finite element method approximations and does not have the problems with boundary effects that the barrier models have (Bolin et al., 2023c).

**1.3 Motivating example**

Accessible, and sustainable transport systems in cities are a core target of 2030 sustainable development goals (SDGs) adopted by the United Nations (UNDP, 2021). Thus, there is an opportunity to apply advanced computational techniques to model the spatial variation in the incidence of road traffic accidents in a linear road network system to aid in accident prevention and multi-disciplinary road safety measures. The motivating example we have used in this paper is ten-years (2010-2019) of daily traffic accident records on the road networks from the central part of Barcelona, Spain. The network is complex enough to motivate a general solution using the proposed non-Euclidean metric on graph model and also compare the results with SPDE network model and barrier model. Further, the study region contains the road segments which observe the maximum daily records of traffic accidents as well as some road segments where there are no records of accidents during the entire study period.

The aim of this paper is two-fold. The principal aim is to apply the new class of Whittle-Matérn fields in the R-INLA framework to traffic accident data. Secondly, to compare and contrast the performance of the proposed model with two other distinct approaches namely, the SPDE network triangulation models and barrier models on linear networks. R (version R 4.3.1) programming language (R Core Team, 2023) has been used for statistical computing and graphical analysis. All computations were conducted on a quad-core Intel i9-4790 (3.60 GHz) processor with 32 GB (DDR3-1333/1600) RAM.

The rest of the paper is organized as follows. Section 2 reports about the study area and data settings along with official sources of data. In Section 3, the first two subsections provide a brief overview of the existing SPDE network triangulation model and barrier model approaches. The final subsection of Section 3 presents Gaussian processes on compact metric graphs introduced in Bolin et al. (2022). Section 4 is devoted to present and compare the model performance along with some related discussions. The paper ends with some concluding remarks in Section 5.

## 2. Data

Barcelona is the largest and capital city of Catalonia, Spain and is located on the northeastern coast of the country. With a population of 1.6 million and a density of 15,748 inhabitants per square km, it is the second most populous municipality in Spain (OpenDataBCN, 2021). The city is a major cultural, economic, and financial center, as well as a transportation hub for southwestern Europe with a well-developed motorway network. In this study, a small area of 4.4 square km in the central part of the city, consisting of 2058 road segments, has been considered as depicted in the left panel of Figure 1 inside the black circle. The road network data has been obtained from the Open Data BCN repository (OpenDataBCN, 2021). The police department in Barcelona keeps records of traffic accidents and related casualties and injuries, which are annually published by Open Data BCN under the *Creative Commons Attribution 4.0* for public sector information. The data is free and available for public sector information.

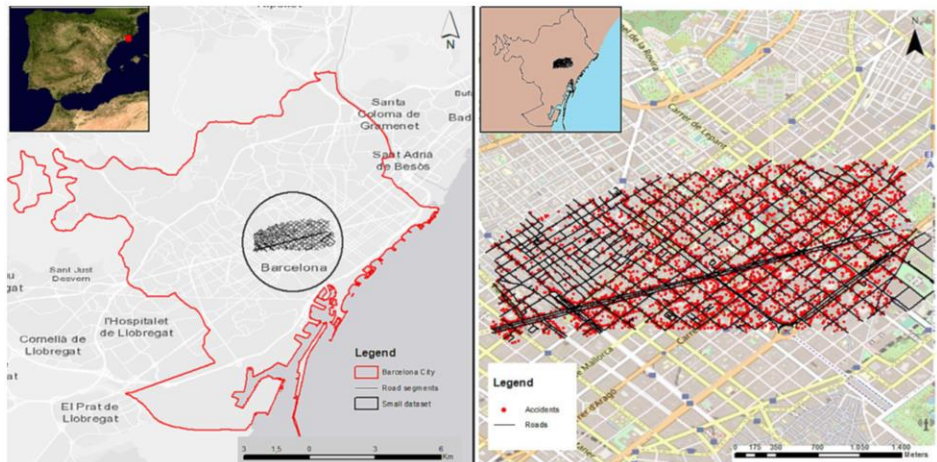

**Figure 1: Geographical location of Barcelona (left) and road network of the study area with traffic accident locations (right)**

During the period from January 2010 to December 2019, there are 11,067 recorded traffic accidents in the study area. The locations of these accidents are shown in red on the road network map in the right panel of Figure 1. The study utilized five datasets from Open Data BCN, which are linked by a record code from 2010 to 2019. The common characteristics recorded in the data consists of a unique event ID, district and neighborhood, postal address and geographical coordinates, and the day and time of occurrence. We included three covariates in our models: road length (ranging from 3.69 to 186.25 meters) with a mean of 81.61 meters, road type (values 1 to 7, with higher values indicating lower traffic), and road speed limit (ranging from 18 to 80 km per hour). Notably, roads with speed limits of 30, 35, and 50 km per hour accounted for 21%, 28%, and 35% of the total sample, respectively. The datasets also include temporal variables such as year, month, and time of the accident. The individual accident locations are adjusted to the nearest road segments. The number of minor injuries has been used as the response variable in the models. Most of the accidents (74.76%) have only one minor injury, followed by two minor injuries (15.42%) and 3 or more minor injuries (3.42%). There were 6.4% of accidents with no minor injuries, and 99.85% of the accidents resulted in no casualties. The number of accidents recorded in each year of the study are similar, with the highest number (1270) in 2016 and the lowest number (847) in 2011. It is worthy to mention that, in the case of network mesh and barrier model the daily minor injuries for individual road segment have been aggregated and included in the centroid of that segment. This means that other temporal covariates related to each accident are not considered in the current study. In contrast, the proposed graph model converts road segments into the edges of a graph and considers accident locations, road network intersections, and the start and end nodes of each road segment as the vertices of the graph. In the first model, the distances to nearby facilities such as bus stops, municipal markets, restaurants, schools, and street markets are calculated from the centroid of each road segment and used as spatial covariates. But in the graph model, these distances are calculated from individual vertices of the graph. A detailed description of generating the vertices and edges of the graph is reported in Section 3.3.

## 3. Methodology

Our discussion in this section initially covers two existing models, namely network mesh and barrier models on linear network, and in the third subsection we explore the recently proposed Whittle-Matérn fields (Bolin et al., 2022) and their application to the selected dataset.

### 3.1 Network triangulation

As discussed in Section 1.2, analysis of spatiotemporal events such as traffic accidents, street crimes, and issues in water and electric connection networks in cities that occur exclusively on linear networks, it has been observed that conventional INLA-SPDE techniques are frequently used to model these events, despite the fact that they are strictly confined to linear networks.

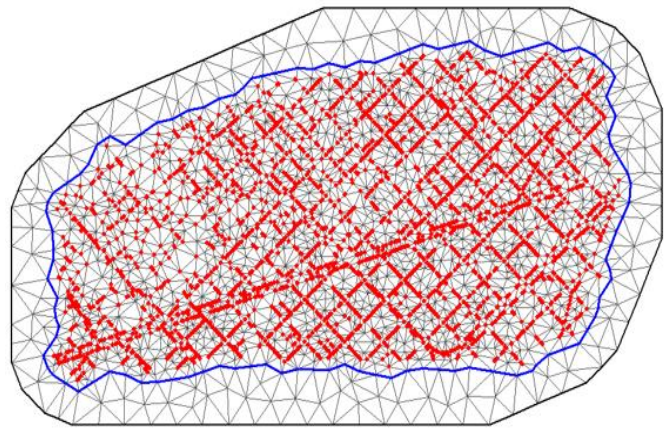

**Figure 2:** Region mesh with non-convex hull boundary in blue and data locations highlighted as red points

When applying the INLA-SPDE method to linear networks, creating a triangulation for the entire region enables fitting of the INLA model in the study area. However, a significant problem arises while predicting events, as the observed events are discrete spatial points located precisely on the road network, whereas models fitted with a region mesh cover the entire study area. This implies that the locations of predicted events can be placed in any area with or without road networks, which is not realistic. Traditional methods of model prediction using a region mesh are, therefore, not appropriate in this context from a scientific perspective.

In the current study, due to close proximity of the road segments, initially a continuous spatial structure is selected for modeling, and triangulation is carried out on the entire study area. In this context, Verdoy (2021) argues that the best mesh for prediction should have a sufficient number of vertices for accuracy but also within a limit to reduce computational time. Following this principle, from a battery of meshes, the best fitted mesh is selected having 2352 vertices. Figure 2 depicts the region SPDE mesh with 11,067 traffic accident locations highlighted as red points. However, the fitted mesh as shown in Figure 2 has a problem when it covers the entire study area. It is unrealistic and ambiguous for the model predictions to cover areas without a road network where traffic accidents are unlikely to occur. This drives the need to design the SPDE triangulation precisely on road networks.

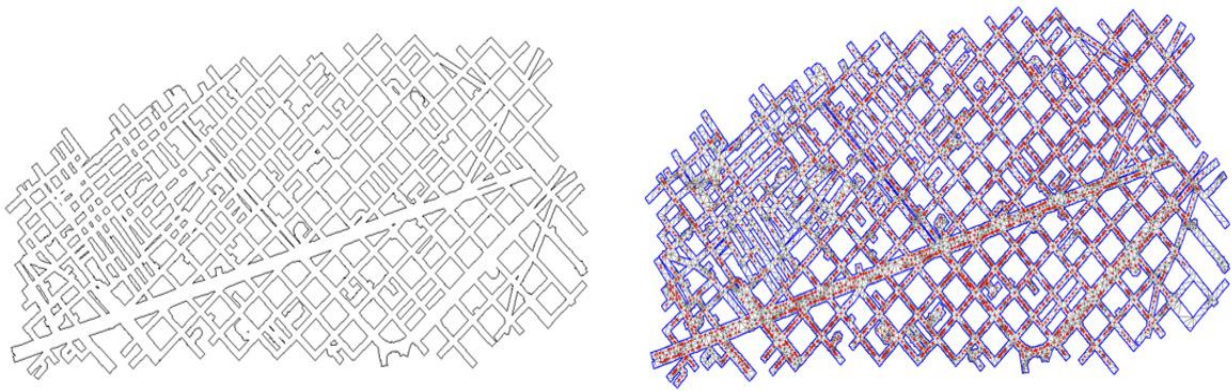

**Figure 3: Buffered road polygon (left) and Network mesh with data locations highlighted in red (right)**

The process consists of three phases: generating a buffer region for each road segment, creating a clipped buffer polygon that covers only the road network, and designing SPDE triangulation on the clipped polygon to form an SPDE network mesh. Choosing the buffer size requires finding a balance between the number of vertices in the triangulated mesh and computational cost (Krainski et al., 2018; Verdoy, 2021). After evaluating various buffer sizes, a 15-meter buffer has been identified to be the best option. The left panel of Figure 3 illustrates the 2058 road segments with a 15-meter buffer around each segment. Following that, we merge individual buffer segments into a single polygon clipped within a bounding box covering the study area. In the final step, we use the centroids of each road segment as the target locations over which we build the initial Delaunay's triangulation.

It is worth to note that, for each road segment, the total number of minor injuries has been aggregated daily and added at corresponding centroids as the response variable. The triangulation is created using the centroids. Figure 3 (right) depicts the SPDE mesh precisely designed on the road network, with accident locations highlighted in red. We report the number of vertices in the network mesh is 14,368. By aggregating data from locations and converting it into event counts per segment, we can utilize Poisson regression models together with a Bayesian approach to model traffic accidents on individual road segments. In fact, we use a spatial Poisson regression method within a Bayesian framework using INLA and SPDE. Recent research conducted in the same study area and utilizing the same dataset, by Chaudhuri et al. (2023b), found that a network mesh model outperformed the SPDE mesh model for the entire study area. Therefore, in this section, we have focused solely on the more efficient network mesh model and compared it with the two other models discussed in Section 3.2 and Section 3.3.

In particular, let $Y_I$ and $E_i$ be the observed and expected number of road traffic accidents on the $i$-th road segment. We assume that conditional on the relative risk, $\rho_i$, the number of observed events follows a Poisson distribution:

$$Y_i \mid \rho_i \sim Po(\lambda_i = E_i \rho_i)$$

where the log-risk is modeled as

$$\log(\rho_i) = \beta_0 + Z_i \beta_i + S(x_i) + \epsilon_I \quad (1)$$

Here, $S(x_i)$ accounts for the spatially structured random effects, and $\epsilon_I$ stands for an unstructured zero mean Gaussian random effect. For the prior selection we have used penalized complexity (PC) priors (Simpson et al., 2017). $Z_I$ represents the spatial covariates. We assigned a vague prior to the vector of coefficients $\beta = (\beta_0, \dots, \beta_p)$ which is a zero mean Gaussian distribution with precision 0.001. All parameters associated to log-precisions are assigned inverse Gamma distributions with parameters equal to 1 and 0.00005. The default prior distributions for all parameters in R-INLA were selected based on commonly used priors in previous studies (Martins et al., 2013; Blangiardo & Cameletti, 2015; Rue et al., 2016; Moraga, 2020). We report that our results are robust against other alternative priors, as we run several cases with different priors obtaining the same results.

To compute the joint posterior distribution of the model parameters, we use an INLA-SPDE method, as introduced by Lindgren et al. (2011). SPDE consists in representing a continuous spatial process, such a Gaussian field (GF), using a discretely indexed spatial random process such as a Gaussian Markov random field (GMRF). In particular, the spatial random process (represented by $S(\cdot)$ explicitly denote dependence on the spatial field, follows a zero-mean Gaussian process with Matérn covariance function (Matérn, 1960) represented as:

$$Cov\left(S(x_i), S(x_j)\right) = \frac{\sigma^2}{2^{\nu-1}\Gamma(\nu)} \left(\kappa \parallel x_i - x_j \parallel\right)^\nu K_\nu\left(\kappa \parallel x_i - x_j \parallel\right)$$

where $K_\nu(.)$ is the modified Bessel function of second kind, $\Gamma$ is a gamma function and $\nu > 0$ and $\kappa > 0$ are the smoothness and scaling parameters, respectively. INLA approach constructs a Matérn SPDE model, with spatial range $r$ and standard deviation parameter $\sigma$.

The parameterized model we follow is of the form:

$$(\kappa^2 - \Delta)^{(\alpha/2)}(\tau S) = W \text{ on } \mathbb{R}^d$$

where $\Delta = \sum_{i=1}^{d} \frac{\partial^2}{\partial x_i^2}$ is the Laplacian, $\alpha = (\nu + d/2)$ is the smoothness parameter, $\tau$ is inversely proportional to $\sigma$, $W$ is Gaussian white noise and $\kappa > 0$ is the scale parameter, related to range $r$, defined as the distance at which the spatial correlation becomes negligible. For each $\nu$, we have $r = \sqrt{8\nu}/\kappa$, with $r$ corresponding to the distance where the spatial correlation is close to 0.1. Note that we have $d = 2$ for a two-dimensional process, and we fix $\nu = 1$, so that $\alpha = 2$ in this case (Blangiardo & Cameletti, 2015). Next, to interpolate discrete event locations to estimate a continuous process in space we have used the SPDE network mesh as depicted in the right panel of Figure 3. The projection matrix is generated using the centroids of individual road segments and triangulations in the mesh. Bakka et al. (2018) suggest that the range value should be determined based on the spatial distribution of events in the study area. In the current study, due to the proximity of accident locations we have decided to use a prior $P(r < 0.01) = 0.01$, meaning that it is highly unlikely the range is less than 10 meters. The parameter $\sigma$ represents the variability of the data and has a prior specified as $P(\sigma > 1) = 0.01$.

### 3.2 Barrier model on linear network

The SPDE triangulations discussed in the previous subsection pose challenges for practical application due to the inclusion of assumptions such as Neumann boundary conditions. While modeling events on complex spatial regions having physical barriers, Bakka et al. (2019) introduced barrier model. As noted in Section 1.2, although the barrier model was not designed specifically for linear road networks, Krainski et al. (2018) applied barrier models in modeling anisotropic behavior, such as the propagation of noise in urban areas. In their study, urban buildings are used as physical barriers and the spatial process is considered on the road network of the study area.

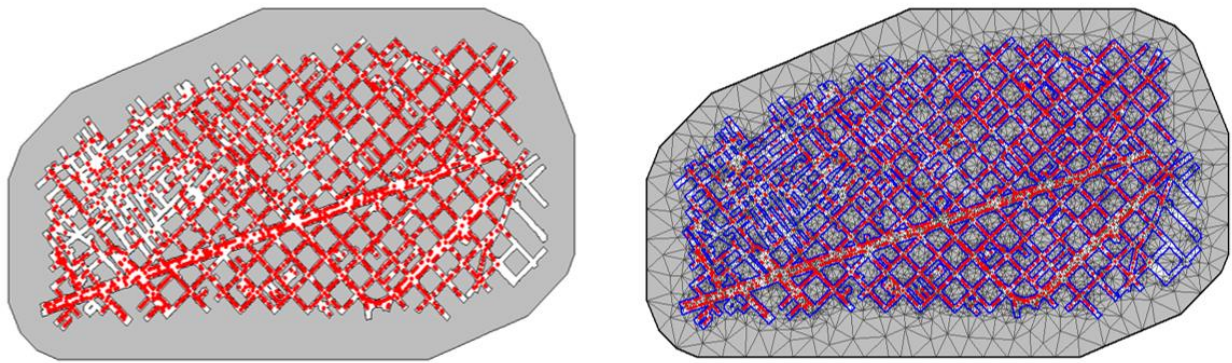

**Figure 4: Barrier object with event locations highlighted as red points (left) and Mesh with barrier object and event locations (right)**

In our current study, we have taken a similar approach to model traffic accidents by utilizing a barrier model in the road network of Barcelona. We have defined polygons of individual road segments with a buffer as our study area and the remaining land areas that do not include roads serve as the physical barriers. The creation of the clipped buffer region and aggregation of the number of minor injuries (which serve as the response variable in the model) at the centroids of each road segment have been accomplished using the same approach outlined in Section 3.1. Similarly, the mesh for the entire study area, as shown in Figure 2, has been constructed using a method similar to that described in Section 3.1. In this case we have used spatial barrier model implemented in recently introduced *INLAspacetime* R package (Krainski et al., 2023). This implementation considers the cgeneric computational approach that is useful to implement new models with INLA. This adaptation leads to a substantial reduction in computation time compared to the original implementation using R-INLA. The *barrierModel.define* function is used to define the model object. The barrier object is depicted in the left panel of Figure 4, where the grey area denotes the physical barrier, and the white area represents the road buffer polygons where spatial dependence is analyzed. Points in red indicate the locations of traffic accidents used as event locations in the model. The triangulation will be created using a barrier model, in which the buffered road polygon serves as the normal terrain and areas without roads serve as physical barriers. The resulting mesh, with the polygon surrounding the barrier (in blue) along with the event locations (in red) are displayed in the right panel of Figure 4.

As mentioned in Section 3.1, our approach involves aggregating event counts at the centroids of individual road segments. By using Poisson regression models and adopting a hierarchical Bayesian spatial model that accounts for barriers, we can model traffic accidents on individual road segments. Our response variable is the aggregate number of minor injuries recorded per day for each individual section of road. Following it, log-risk in Equation 1 can be modified to:

$$\text{Log}(\rho_i) = \beta_0 + Z_i\beta_i + u(s_i) + \epsilon_I \tag{2}$$

Here, $\beta_0$ corresponds to the intercept, $Z_I$ represents the spatial covariates mentioned in Section 2 and $\epsilon_I$ stands for an unstructured zero mean Gaussian random effect and log Gamma precision parameters 0.5 and 0.01, defined as PC priors (Simpson et al., 2017). We assigned default priors for all fixed-effect parameters to minimize their impact on the posterior distribution. $u(s)$ is a non-stationary spatial random effect. Bakka et al. (2019) in their proposal suggested using a finite element method which is based on the SPDE approach. The proposed method involves a system of two SPDEs, where one is applied to the barrier region and the other to the rest of the area. The system of stochastic differential equations in question has a solution that exhibits a non-stationary spatial effect, represented as $u(s)$. The system can be mathematically modeled as a set of stochastic differential equations, which provide a continuous weak solution to the estimation problem:

$$u - \nabla \cdot \frac{r_b^2}{8} \nabla u = r_b \sqrt{\frac{\pi}{2}} \sigma_u W, \text{ on } \Omega_b \tag{3}$$

and

$$u - \nabla \cdot \frac{r^2}{8} \nabla u = r \sqrt{\frac{\pi}{2}} \sigma_u W, \text{ on } \Omega_n \tag{4}$$

where $u(s)$ is the spatial effect, $\Omega_b$ the barrier area and $\Omega_n$ is the remaining area and their disjoint union gives the whole study area $\Omega$. Ranges for the barrier and remaining areas are represented by $r$ and $r_b$ respectively. $\Sigma_u$ is the marginal standard deviation. $\nabla$ is equal to $\left(\frac{\partial}{\partial x}, \frac{\partial}{\partial y}\right)$ and $W$ is Gaussian white noise. In contrast to stationary spatial effects, this method implies the creation of a GMRF at a local level, consisting of two governing equations, one for the normal area (buffered road polygon) and the other for the barrier area (areas without roads). The spatial effect prior is determined by two unknown hyperparameters, namely the standard deviation ($\sigma_u$) and the range in the normal area ($r$), while the range in the barrier area ($r_b$) is maintained at a fixed, low value.

Therefore, the system in Equation 3 and 4 represents a form of local averaging, with dependence on nearby values. This approach ensures that when two points are separated by physical barriers, the small range in the barrier area prevents local averaging, forcing dependency to focus on movement around the barrier through local averaging in the buffered road polygon area. The system of differential equations in Equation 3 and 4 can be solved by constructing a Delaunay triangulation of the study area (as shown in Figure 4) and applying the finite element method, as described in Bakka et al. (2019). For the two hyperparameters in the model that define the covariance structure of $u(s)$, PC priors were assigned following the parametrization outlined in Simpson et al. (2017) and Fuglstad et al. (2019).

### 3.3 Whittle-Matérn fields on linear networks

Previous subsection on explicit network triangulation and ongoing research on barrier models for complex land structures have highlighted issues related to boundary effects, including the creation of artifact spatial dependencies on the boundary. In standard meshes, boundaries are typically outside the spatial domain of interest, allowing for identification and elimination of these dependencies. However, in more complex meshes like network triangulation or barrier models, boundaries lie within the spatial domain, making it challenging to identify and eliminate these dependencies. A comprehensive analysis of boundary effects is presented in Section 4. However, a different approximation is needed in which the INLA-SPDE approximation does not cause these fictitious spatial dependencies.

Literature shows statistical models are required to be defined on linear networks, such as connected river or street networks (Baddeley et al., 2017; Cronie et al., 2020). In such cases, it is necessary to define a model using a metric on the network rather than the Euclidean distance between points. However, constructing Gaussian fields over linear networks, or more generally on metric graphs, presents a challenge. This is due to the difficulty of finding flexible classes of functions that are positive definite when a non-Euclidean metric is used (Bolin et al., 2023a; Bolin et al., 2023b). In a specific type of metric graph with Euclidean edges, Anderes et al. (2020) demonstrated that, for graphs with Euclidean edges, it is possible to define a valid Gaussian field by using a Matérn covariance function:

$$r(s,t) = \frac{\sigma^2}{2^{\nu-1}\Gamma(\nu)} \big(\kappa d(s,t)\big)^{\nu} K_{\nu}\big(\kappa d(s,t)\big) \qquad (5)$$

where $d(\cdot,\cdot)$ is the resistance metric and $0 < \nu \leq 1/2$. The limitation of $\nu \leq 1/2$ means that we cannot use this approach to create differentiable Gaussian processes on metric graphs, even if they have Euclidean edges (Bolin et al., 2022). In their recent work, Bolin et al. (2023b) demonstrated that no Gaussian random fields exist on general metric graphs that possess both isotropic and Markov properties. Due to this constraint, as well as some other complex challenges in creating Gaussian fields via covariance functions on non-Euclidean spaces, Bolin et al. (2022) take a different approach and focus on creating a Gaussian random field $u$ on a compact metric graph $\Gamma$ as a solution to a SPDE

$$(\kappa^2 - \Delta)^{\alpha/2}(\tau u) = \mathcal{W}, on\ \Gamma \qquad (6)$$

where $\alpha = \nu + 1/2$, $\Delta$ is the Laplacian equipped with suitable boundary conditions in the vertices, and $\mathcal{W}$ is Gaussian white noise (Bolin et al., 2023c). The proposed models, known as **Whittle-Matérn** fields, represent a logical progression from Gaussian fields with Matern covariance functions in Euclidean spaces to the more complex non-Euclidean metric graph environment (Bolin et al., 2022; Bolin et al., 2023c). Bolin et al. (2023c) have shown in their proposed methodology that for $\alpha \in \mathbb{N}$, when these fields exhibit Markov properties, it is possible to conduct exact likelihood-based inference and spatial prediction with high computational efficiency. This progress significantly enhances the practicality of employing **Whittle-Matérn** fields in statistical applications, even when working with large datasets, eliminating the requirement for approximations such as finite element method. For a detailed description of the methodology, please refer to the works of (Bolin et al., 2022; Bolin et al., 2023b; Bolin et al., 2023c).

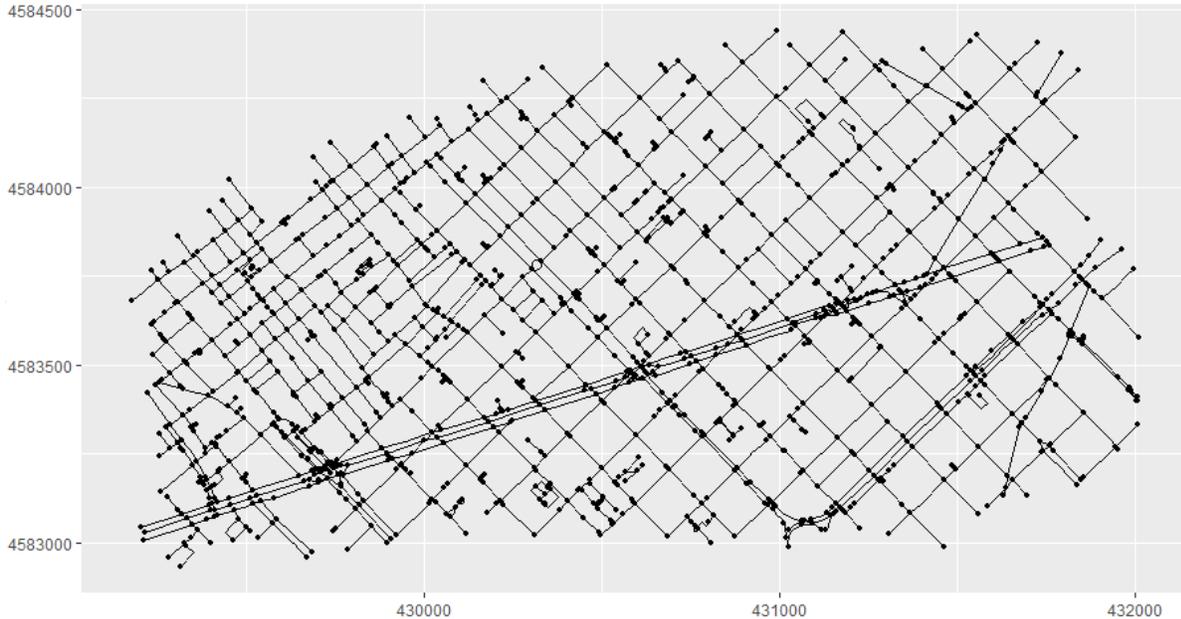

**Figure 5: Metric graph structure generated using road networks of the study area**

In the current study, we have implemented the proposed methodology using the recently introduced R package *MetricGraph* (Bolin et al., 2023d). We have used the same traffic accident dataset of Barcelona city spanning from January 2010 to December 2019 reported in Section 2. We focused on the number of minor injuries as the response variable for our modeling process. To employ a graph model, we first converted the dataset into a graph data structure that is compatible with the model. In this structure, we have represented individual accident locations, start and end points of road segments, and intersecting points of road segments as nodes or vertices, while the connecting road segments for the nodes are represented as edges. Initially, we employed the function *metric_grap$new* on the road network shape files containing 2058 road segments to generate the metric graph object. Figure 5 illustrates the resulting metric graph,

consisting of 1288 vertices and 2050 edges. In the subsequent phase, to incorporate the traffic accident data into the metric graph, we utilized the function *add_observations*.

During this procedure, we included all the relevant covariates and the response variable (minor injuries) at the exact locations of the traffic accidents along the road network. Figure 6 displays the metric graph with 11067 individual traffic accident locations along with their respective counts of minor injuries, represented by a color scale. As outlined in Section 2, the majority, approximately 74.76%, of minor injuries have a value of 1, ranging from a minimum of 0 to a maximum of 12.

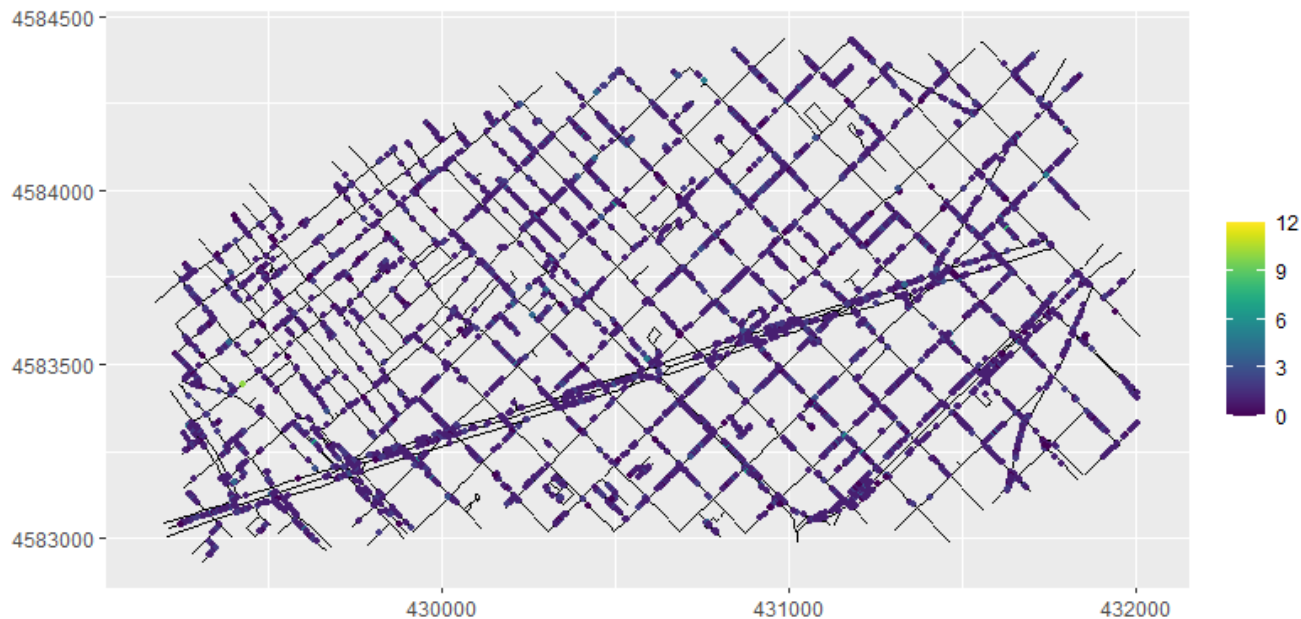

**Figure 6: Metric graph with observations of the traffic accident locations as nodes and road networks as edges**

It is important to note that we have aggregated the number of minor injuries for different time instances at the same location and considered it as a single vertex. Thus, in this model and also in the previous two models (mentioned in Section 3.1 and Section 3.2) no temporal covariates are considered. In the following step, we generated the INLA model object using the *graph_spde* function. Subsequently, we created the data object employing the *graph_data_spde* function. In the final step, utilizing the R-INLA framework, we constructed the formula object and set up the *inla.stack* object. Consistent with Section 3.1, we assume that given the relative risk, $\rho_i$, the observed event count follows a Poisson distribution, as described in Equation 1. Finally, we fitted the model within the R-INLA framework.

## 4. Results and discussion

In this section, we present the results of the comparative analysis of the three methodologies outlined in Section 3. The evaluation was conducted using the same dataset, allowing for a direct comparison of the performance of these distinct modeling approaches. It is essential to highlight that temporal covariates are not incorporated into any of the modeling procedures. For both network mesh model and barrier model, we executed batteries of similar models based on the argument values to create SPDE triangulation. The default prior distributions for all parameters in R-INLA are selected based on commonly used priors in previous studies (Rue et al., 2017; Moraga, 2020). Our results indicate that our findings are robust against alternative priors, as we ran several cases with different priors and obtained the same results. In the case of the metric graph model, we executed several models with different log-prior probability density for the model parameters and ultimately selected the best fitted model. We assessed the performance of the models from the three different approaches using deviance information criterion (DIC) and the Watanabe–Akaike information criterion (WAIC), balancing model accuracy against complexity (Spiegelhalter et al., 2002). We have used conditional predictive ordinate (CPO) value (Gelfand et al., 1992) which also acts as a selection measure; smaller value of CPO indicates a better prediction quality of the model. Execution time for each modeling approach has also been reported as a measure of comparison. In Table 1, we report the selected models with the lowest DIC, WAIC, CPO, and execution time for each of the three categories from their respective battery of models.

Table 1: DIC, WAIC and CPO values of Fitted Models

|  | Network Mesh Model | Barrier Model (INLAspacetime) | Metric Graph Model |
|---|---|---|---|
| **DIC** | 23390.17 | 23364.32 | ***16059.86*** |
| **WAIC** | 23382.56 | 23353.28 | ***16051.75*** |
| **CPO** | 0.3252899 | ***0.3252616*** | 0.3257345 |
| **Execution Time (Secs.)** | 21.2 | 16.1 | ***9.6*** |

Analyzing the DIC values, it is evident that the metric graph model stands out with the lowest DIC value (16059.86), while the network mesh and barrier models are similar with DIC values of 23390.17 and 23364.32, respectively. A lower DIC value indicates a better fit, so we can conclude that the metric graph model is the best-fitted model among the three. Similarly, when examining the WAIC values, we noticed that the metric graph model has the lowest value (16051.75), again indicating that it is the best-fitted model. The difference between the WAIC values of the other two models is very small, suggesting that they have very similar performance in terms of fitting the data. Finally, the CPO values for all three models are close to each other, indicating that they all have similar predictive performance. Additionally, it is important to note that the metric graph model has the shortest execution time among the three models, taking only 9.6 seconds. The network mesh model, on the other hand, requires the longest time at 21.2 seconds. The barrier model falls in between, taking 16.1 seconds. This indicates that the metric graph model is not only more accurate but also more efficient in terms of computation compared to the other two models. In summary, the metric graph model has the best performance according to both DIC and WAIC, while all three models have similar predictive performance according to CPO. These results suggest that the metric graph model is the best model among the three fitted models for describing the data as well as computational efficiency.

While comparing the significance of the fixed effects for the three modeling techniques, we have observed that the covariates included in all models do not exhibit a statistically significant influence on the outcome. Moreover, the Appendix includes marginal posterior distributions of model hyperparameters for three different models: network mesh model in Figure 7, barrier model in Figure 8, and metric graph model in Figure 9. The spatial range values for each model are reported as follows: 0.0523 Km (equivalent to 52.3 meters) for the network mesh model, 0.0183 Km (equivalent to 18.3 meters) for the barrier model, and 0.0648 Km (equivalent to 64.8 meters) for the graph model.

It is worth mentioning that, while modeling random spatial events on linear networks by using SPDE triangulation precisely on linear network, it is important to carefully consider the potential limitations and trade-offs associated with this approach. The barrier model, while generally easy to implement and adept at handling sparse data and complex physical barriers, was not originally intended for application to linear networks. Despite this, our study explores the use of the barrier model as an alternative methodology in this context. However, while using these two techniques for modeling spatial relationships have significant limitation in the form of boundary effects. These effects can result in biased estimates and prediction errors, particularly in proximity to the boundary, if the mesh does not cover the entire domain. As a general rule, the variance near the boundary is inflated by a factor of two along straight boundaries and by a factor of four near right-angled corners (Lindgren et al., 2011; Lindgren & Rue, 2015). The complex boundary region of the buffer road network with several right-angled corners makes the process critical. Thus, these two methods are not a good fit to handle spatial random events on linear networks. The model results in the current study also establish this fact.

On the other hand, the proposed metric graph methodology by Bolin et al. (2022) has two significant advantages. Firstly, it has Markov properties if $\alpha \in \mathbb{N}$, implying that the precision matrices of the finite dimensional distributions of the process will be sparse, which simplifies the use of the model for big datasets without the need for finite element methods or other approximations. Secondly, the model is well-defined for any compact metric graph, not just the subclass with Euclidean edges. This approach extends Gaussian fields with Matérn covariance functions on Euclidean domains to non-Euclidean metric graph settings (Bolin et al., 2022). Additionally, the recent introduction of the R package *MetricGraph* enables the efficient generation and manipulation of metric graphs (Bolin et al., 2023d). It further simplifies operations and visualizations of data on these graphs, as well as the generation of a diverse set of random fields and SPDE in these spaces. Importantly, the integration of R packages *INLA* and *inlabru* facilitates the application of Bayesian statistical models on metric graphs.

## 5. Conclusion

The current study discusses the challenges of using INLA and traditional SPDE method in implementing Bayesian spatiotemporal modeling in complex linear networks. These challenges call for a comprehensive and novel approach to address them. This may involve improving the SPDE triangulation approach, especially for linear networks, or developing a generalized methodology to model spatial and spatiotemporal events in complex land structures. In this context the current study initially explored the innovative and realistic computational strategy for constructing spatial triangulations constrained to linear network topologies. But the presence of complex boundary regions resulted in artificial spatial structures and dependencies leading to unavoidable boundary effects in the model. To address this, as an alternative computational strategy, nonstationary barrier models have been implemented. It is important to note that the barrier model was not originally designed for linear networks; it is more suitable for spatial regions with physical barriers. But in the current study we have explored it as a nonstationary alternative. However, as the boundaries of the complex road networks remained within the spatial domain of interest it is challenging to reduce these boundary effects. Therefore, the results we obtained using the barrier model approach are also not very promising. Finally, the current study explored the recently introduced new class of Gaussian processes on compact metric graphs, the Whittle-Matérn fields defined by a fractional SPDE on a metric graph. This approach extends a well-defined framework for any compact metric graph. While comparing model performance using evaluation metrics, we observed that the metric graph model outperforms network triangulation and barrier models.

Currently, the methodologies we have discussed utilize models that incorporate spatial covariates only; no temporal covariates have been integrated. As a result, the capability to predict future events is suboptimal, particularly in real-time scenarios where temporal covariates can play a vital role. We are actively exploring the incorporation of temporal covariates using the recently introduced *INLAspacetime* R package for the network mesh and barrier models, along with the *MetricGraph* R package for the metric graph model. The objective is to enhance the model performance in the spatiotemporal context, thereby motivating new possibilities for investigation in the field of Bayesian spatiotemporal modeling for linear networks. Additionally, applying the flexible, robust and computationally efficient metric graph methodology can enhance prediction accuracy. This, in turn, can lead to a wide array of improved applications and more effective management and control of real-time urban and environmental challenges, particularly those involving linear networks like street or river systems.

**Conflict of interest**

The authors declare that they have no conflict of interest.

**Ethical statement**

Author and co-authors testify that, this manuscript is original, has not been published before and is not currently being considered for publication elsewhere. We know of no conflicts of interest associated with this publication and there has been no financial support for this work.

**Authors' contributions**

MS and PJ had the original idea for the paper. MS, PJ and SC designed the study. The bibliographic search and the writing of the introduction were carried out by all the authors. Data collection and cleaning were performed by SC and DV. The methods and statistical analysis were chosen and performed by SC and PJ and SC created the tables and figures. All authors wrote the results and the discussion. The writing and final editing was done by all authors. All authors reviewed and approved the manuscript.

# Appendix

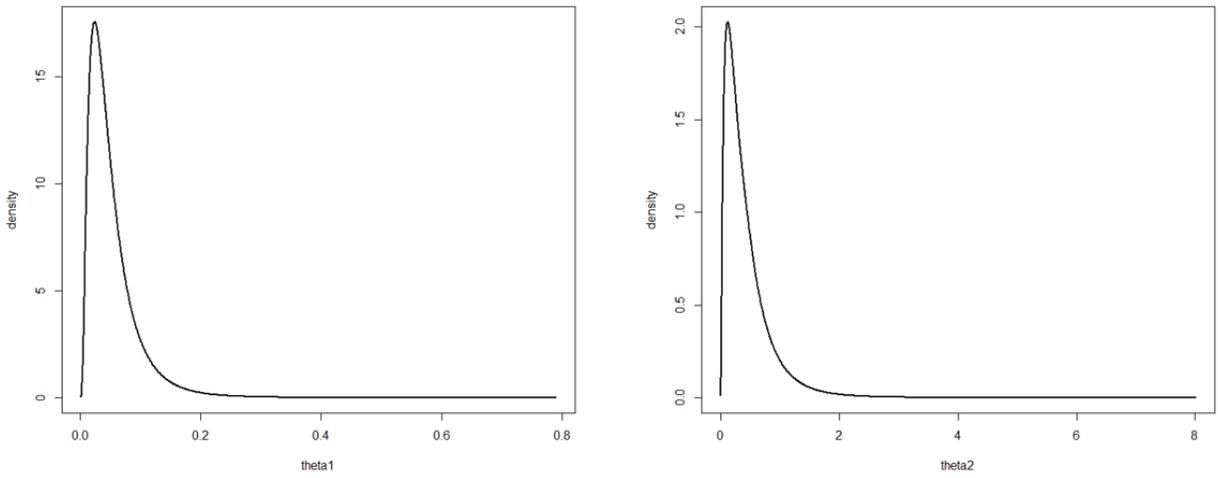

**Figure 7: Marginal posterior distributions of network mesh model hyperparameters: $\theta_1$ (left) and $\theta_2$ (right)**

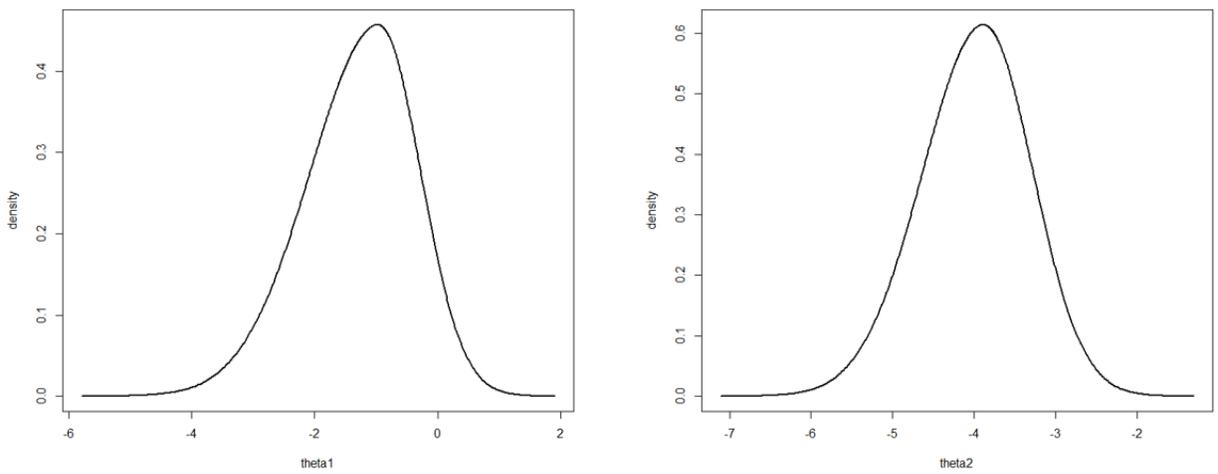

**Figure 8: Marginal posterior distributions of barrier model hyperparameters: $\theta_1$ (left) and $\theta_2$ (right)**

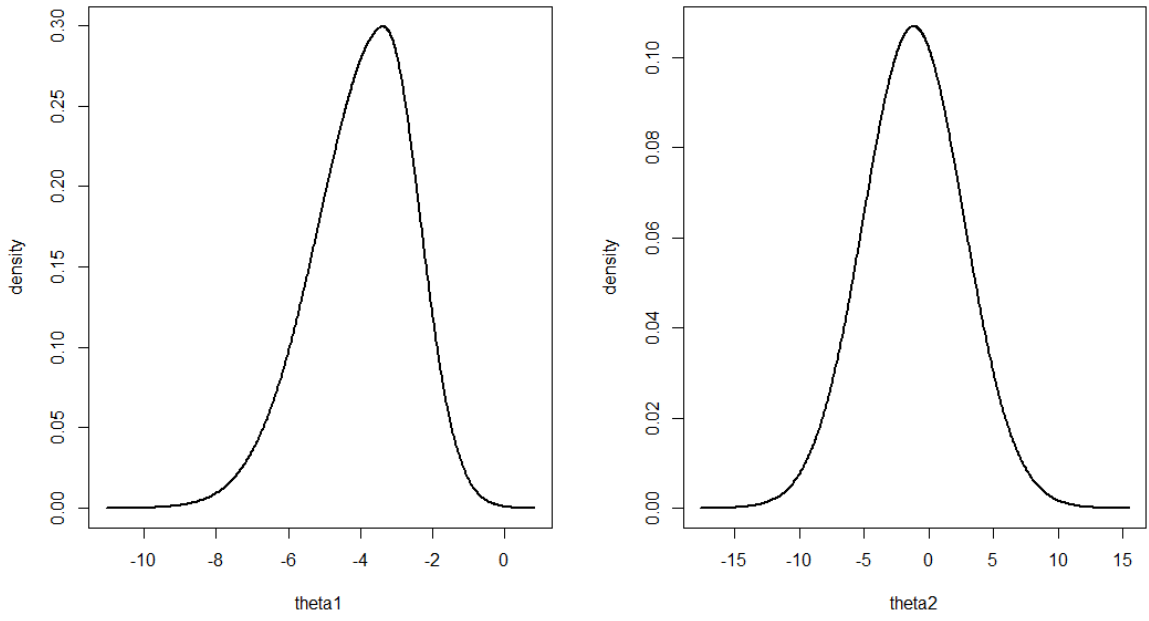

**Figure 9: Marginal posterior distributions of graph model hyperparameters: $\theta_1$ (left) and $\theta_2$ (right)**